\documentstyle[prd,aps]{revtex}
\newcommand{\be}{\begin{equation}}
\newcommand{\ee}{\end{equation}}
\newcommand{\bea}{\begin{eqnarray}}
\newcommand{\eea}{\end{eqnarray}}
\def\aprle{\buildrel < \over {_{\sim}}}
\def\aprge{\buildrel > \over {_{\sim}}}
\begin{document}
\draft
\input epsf
\twocolumn[\hsize\textwidth\columnwidth\hsize\csname
@twocolumnfalse\endcsname

\title{Vacuum energy and relativistic invariance}
\author{E. Kh. Akhmedov${}^{(a,b)}$ 
}
\address{$^{(a)}${\em
CFIF, Departamento de Fisica, Instituto Superior T\'ecnico,
P-1049-001 Lisboa, Portugal}} 
\address{${}^{(b)}${\em National Research
Centre Kurchatov Institute, Moscow 123182, Russia}} 
\date{v1: April 4, 2002; 
v2: June 4, 2002}
\maketitle
\begin{abstract}
It is argued that the zero-point energies of free quantum fields diverge 
at most quadratically and not quartically, as is generally believed. 
This is a consequence of the relativistic invariance which requires 
that the energy density of the vacuum $\rho$ and its pressure $p$ satisfy 
$\rho=-p$. The usually obtained quartic divergence is an artifact 
of the use of a noninvariant regularization which violates this 
relation. One consequence of our results is that the zero-point energies 
of free massless fields vanish. Implications for the cosmological 
constant problem are briefly discussed. 
\end{abstract}
\vskip1.2pc ]

Vacuum energy is an additive part of the energy of any physical system. 
It plays no role in non-gravitational physics where only energy differences 
are observable; however, it is of profound importance for gravity. 
The reason for this is that the energy itself (and not energy differences)  
is the source of gravity.  The vacuum energy enters into the Einstein 
equation as a cosmological constant term. 
It has long been known from cosmological observations that the energy 
density of the vacuum $\rho$ must be smaller than the critical density of 
the universe, $\rho_c\sim 10^{29}$ g/cm$^3\simeq 4.3\times10^{-47}$ GeV$^4$; 
recent observations \cite{ct} indicate that $\rho\ne 0 $ and is actually 
about $0.7 \rho_c$. At the same time, it is frequently stated that the 
natural value for the vacuum energy density is $\sim M_{\rm Pl}^4$ which is 
more than 120 orders of magnitude larger. This huge discrepancy between
the observed and ``natural'' values of the vacuum energy constitutes the
so called cosmological constant problem \cite{rev}. 

In the present Letter we argue that the above estimate of the natural value 
of the vacuum energy is actually incorrect and results from the use of a  
regularization of the zero-point energies of quantum fields that does not 
respect the relativistic invariance of the problem. When invariant 
regularizations are employed, the usual quartic divergences of the 
zero-point energies, which lead to the above predictions, do not arise, 
and these energies diverge at most quadratically. In particular, in flat 
\mbox{spacetimes} the zero-point energies of free massless fields vanish. 
While our observations do not solve the cosmological constant problem, 
we believe they may shed some new light on it. 

In general, the vacuum energy gets contributions from both classical and 
quantum effects.  Classical contributions may arise from non-vanishing 
effective potentials of scalar fields or scalar combinations of higher-spin 
fields; in particular, cosmological phase transitions may lead to ``latent 
heat'' contributions to $\rho$. Quantum contributions are those of the 
zero-point energies of all quantum fields. It is these quantum contributions 
to the vacuum energy that we shall be concerned with here. 


Consider the zero-point energy of a free real scalar field $\phi$ of mass 
$m$ (the generalization to the cases of a complex scalar field or
higher-spin fields is straightforward). 
Inserting into the energy-momentum tensor $T_{\mu\nu}$ of the field $\phi$ 
the plane-wave decomposition of this field and taking the vacuum expectation
value of the component $T_{00}$, one arrives at the following well-known 
result for the contribution of the field $\phi$ to the energy density of 
the vacuum:   
\be
\rho=\frac{1}{2}\int\frac{d^3k}{(2\pi)^3}\sqrt{{\bf k}^2+m^2}\,.
\label{rho1}
\ee
This is just the sum over all modes of the zero-point energies $\omega_{k}
/2=\sqrt{{\bf k}^2+m^2}/2$. The integral in (\ref{rho1}) is clearly divergent; 
introducing a 3-dimensional ultraviolet momentum cutoff $\Lambda$ one 
finds that the integral diverges as $\Lambda^4$. This is the usual quartic 
divergence of the zero-point energies discussed in many quantum field theory 
textbooks. Taking $\Lambda$ to be the highest known energy scale, the Planck 
scale, one arrives at the estimate $\rho\sim M_{\rm Pl}^4$ mentioned above.

To see what the flaw of this argument is, let us consider the contribution 
of the zero-point fluctuations of the field $\phi$ to the vacuum pressure 
$p$. First, let us recall that, quite generally, on the grounds of 
relativistic invariance, the energy-momentum tensor of the vacuum must be
of the form 
\be
T_{\mu\nu}^{vac}= const\cdot g_{\mu\nu}\,,
\label{vac1}
\ee 
where $g_{\mu\nu}$ is the spacetime metric.  
The vacuum pressure is given by the components $T_{ii}^{vac}$ ($i=1, 2, 3)$ 
of this tensor. From eq. (\ref{vac1}) it then immediately follows that the 
vacuum energy density $\rho$ and pressure $p$ must satisfy \cite{f1}
\be
\rho=-p\,.
\label{rel1}
\ee
Next, we calculate the contribution of the field $\phi$ to the vacuum 
pressure. Following the same steps that led to (\ref{rho1}) one finds   
\be
p=\frac{1}{6}\int\frac{d^3k}{(2\pi)^3}\frac{{\bf k}^2}{\sqrt{{\bf k}^2+m^2}}\,,
\label{press1}
\ee
where the isotropy of the vacuum has been used. It should be noted that the 
derivations of eqs. (\ref{rho1}) and (\ref{press1}) were actually carried out 
in Minkowski spacetime; we shall comment on the general case of curved 
spacetimes later on. 
  
We now want to see if the relativistic invariance condition (\ref{rel1}) 
is satisfied by expressions (\ref{rho1}) and (\ref{press1}). Since the 
corresponding integrals diverge, they can only be compared after a  
regularization procedure has been applied. 
Any regularization includes, explicitly or implicitly, some subtraction; 
our point here is that the subtraction must be performed in such a way
that the relativistic invariance condition (\ref{rel1}) be satisfied. 

We first check the standard regularization of the integrals in (\ref{rho1}) 
and (\ref{press1}) by a 3-dimensional momentum cutoff $\Lambda$. 
One can write 
$\rho=(1/4\pi^2)I_1$, $p=(1/12\pi^2) I_2$, 
where 
\be
I_1=\int_0^\Lambda dx \, x^2 \sqrt{x^2+m^2}\,,\quad 
I_2=\int_0^\Lambda dx \frac{x^4}{\sqrt{x^2+m^2}}\,.
\label{I1I2}
\ee
The relativistic invariance condition (\ref{rel1}) 
implies
\be
I_2=-3 I_1\,.
\label{cond1}
\ee
Since both $I_1$ and $I_2$ given in eq. (\ref{I1I2}) are positive 
definite, this condition is clearly violated.
To see what kind of terms violate it, let us calculate the 
integral $I_2$ by parts. This gives
\be
I_2=\Lambda^3\sqrt{\Lambda^2+m^2}-3 I_1\,.
\label{viol}
\ee
One sees that eq. (\ref{cond1}) is violated by the first (off-integral)
term on the r.h.s. of (\ref{viol}), which diverges as $\Lambda^4$
in the limit $\Lambda\to \infty$. 
Direct calculation of the integrals $I_1$ and $I_2$ yields 
\bea
3I_1=\frac{3}{4}\left[\Lambda(\Lambda^2+m^2)^{3/2}-\frac{1}{2} \Lambda m^2
\sqrt{\Lambda^2+m^2} 
\right. 
\nonumber \\
\left. 
-\frac{1}{2} m^4 \ln\left(\frac{\Lambda+
\sqrt{\Lambda^2+m^2}}{m}\right)\right]\,,
\label{I1}
\eea
\bea
I_2=\frac{3}{4}\left[\frac{1}{3}\Lambda^3\sqrt{\Lambda^2+m^2}-\frac{1}{2} 
\Lambda m^2 \sqrt{\Lambda^2+m^2}
\right. 
\nonumber \\
\left. 
+\frac{1}{2} m^4 \ln\left(\frac{\Lambda+
\sqrt{\Lambda^2+m^2}}{m}\right)\right]\,.
\label{I2}
\eea
We observe that, while the terms $\sim m^4\ln(\Lambda)$ in these
expressions satisfy the relativistic invariance condition (\ref{cond1}),
the terms of order $\Lambda^4$ and $\Lambda^2 m^2$ do not. Similar results 
are obtained if, instead of using a sharp cutoff, one employs smooth
cutoffs by introducing some convergence factors 
into the integrands of $I_1$ and $I_2$ in eq. (\ref{I1I2}) and extending
the integration to infinity. 
The reason for the violation of condition 
(\ref{cond1}) can be readily understood: The 3-dimensional momentum cutoffs 
explicitly violate relativistic invariance. Thus, we have encountered a well 
known problem when the regularization used to make sense out of infinite 
expressions does not respect a symmetry of the problem. In this situation 
one may adopt one of the two different approaches: (1) Continue using the 
noninvariant regularization but introduce (noninvariant) counter terms which 
would compensate for this and restore the invariance; (2) Use an invariant 
regularization. The first approach turns out to be not very useful for
the problem under consideration. The reason is that  
the requirement of the relativistic invariance does not fix the counter 
terms uniquely as one is always free to add an arbitrary constant to the 
Lagrangian; this would modify $T_{\mu\nu}^{vac}$ by a term of the form 
(\ref{vac1}). We therefore concentrate on the second approach. 

{\em Dimensional regularization.} In a $d$-dimensional spacetime with one
time and $d-1$ space dimensions one has to replace eqs. (\ref{rho1}) and 
(\ref{press1}) by 
\bea
\rho=\frac{1}{2}\mu^{4-d}\int\frac{d^{d-1}k}{(2\pi)^{d-1}}
\sqrt{{\bf k}^2+m^2}
\nonumber \\
=\mu^4\frac{\Gamma(-d/2)}{2(4\pi)^{(d-1)/2}
\Gamma(-1/2)}\left(\frac{m^2}{\mu^2}\right)^{d/2}\,,
\label{rho2}
\eea
\bea
p=\frac{1}{2(d-1)}\mu^{4-d}\int\frac{d^{d-1}k}{(2\pi)^{d-1}}\frac{{\bf k}^2}
{\sqrt{{\bf k}^2+m^2}}
\nonumber \\
=\mu^4\frac{\Gamma(-d/2)}{4(4\pi)^{(d-1)/2}\Gamma(1/2)}\left( 
\frac{m^2}{\mu^2}\right)^{d/2}\,,
\label{press2}
\eea
where $\mu$ is a parameter with dimensions of mass introduced so as to keep 
the correct dimensions of $\rho$ and $p$. From the well-known property 
of the $\Gamma$ function $\Gamma(-1/2)=-2\Gamma(1/2)$ it immediately follows 
that eqs. (\ref{rho2}) and (\ref{press2}) satisfy the relativistic invariance 
condition (\ref{rel1}). This had to be expected since dimensional 
regularization respects relativistic invariance.  
Setting $d=4-\epsilon$ one obtains in the limit $\epsilon\to 0$ 
\bea
\rho=-p=-\frac{m^4}{64\pi^2}\left(\frac{2}{\epsilon}+\frac{3}{2}
-\gamma-\ln(m^2/4\pi\mu^2)\right)
\nonumber \\
\equiv
-\frac{m^4}{64\pi^2}\left(\ln(\Lambda^2/m^2)
+\frac{3}{2}\right)\,,
\label{dimr}
\eea
where $\gamma\simeq 0.5772$ is the Euler-Mascheroni constant, and in the last 
equality we traded the parameter $\mu$ for a mass scale $\Lambda$ using the 
$\overline{\rm MS}$ renormalization scheme convention. Direct inspection shows 
that the coefficients of the logarithmically divergent terms in (\ref{dimr}) 
coincide with those obtained using the 3-momentum cutoff.   

It should be noted that dimensional regularization is not suitable for 
studying the power-law divergences \mbox{since} the integrals of the type
$\int
d^d k (k^2)^\alpha$ vanish identically in this regularization scheme. 
What we actually learned from the above 
calculation is that the logarithmically divergent terms in $\rho$ and $p$ 
indeed satisfy the relativistic invariance condition (\ref{rel1}) and that 
the naive 3-momentum cutoff calculation produced them with correct 
coefficients. 

{\em Four-momentum cutoff}. In order to be able to use a relativistically 
invariant 4-momentum cutoff, one needs manifestly covariant expressions 
for $\rho$ and $p$. One possibility would be to try to convert the integrals 
over 3-momentum in eqs. (\ref{rho1}) and (\ref{press1}) into 4-momentum
integrals. However, this procedure is not well defined because the integrals 
are divergent. To see what kind of difficulties can arise along this route, 
let us assume that the integrals in (\ref{rho1}) and (\ref{press1}) are 
regulated, and that the regularization (i) either does not modify the 
integrands or introduces into them an $m$-independent convergence factor, 
and (ii) does not affect the covariance properties of the integrals. Under 
these conditions, one can differentiate the integrals with respect to $m^2$. 
From (\ref{rho1}) one then finds
\be
\frac{\partial \rho}{\partial m^2}=\frac{1}{2}\int\frac{d^3k}{(2\pi)^3 
2\omega_{k}}
\,.
\label{drho1}
\ee
The integral on the r.h.s. is well known to be relativistically
invariant; it can be written in the manifestly invariant form as 
\be
\frac{\partial \rho}{\partial m^2}=\frac{i}{2}\int\frac{d^4k}{(2\pi)^4}
\frac{1}{k^2-m^2+i\varepsilon}=\frac{1}{2} D_F(x,x)\,,
\label{drho2}
\ee
where $D_F(x,y)$ is the coordinate-space Feynman propagator of the field 
$\phi$. 
Integrating eq. (\ref{drho2}) one finds, up to an $m$-independent
constant, 
\be
\rho=
\frac{1}{2}\int\frac{d^4 k_E}{(2\pi)^4}\ln\left(1+\frac{m^2}{k_E^2}\right)\,,
\label{rho3}
\ee
where the integral was Wick-rotated to Euclidean space, so that the 
4-momentum cutoff can now be utilized. With this cutoff in place, one
finds from (\ref{rho3}) 
\be
\rho=\frac{1}{64\pi^2}\left[\Lambda^4\ln\left(\frac{\Lambda^2+m^2}{\Lambda^2}
\right)+\Lambda^2 m^2
-m^4\ln\left(\frac{\Lambda^2+m^2}{m^2}\right)\right] 
\label{rho4}
\ee
The obtained result appears to be
quite sensible: it is just the 1-loop contribution $V_1$ to the effective 
potential $V(\phi)$ of a free scalar field. The mass term in the Lagrangian 
of $\phi$ can be considered as a tree-level potential $U(\phi)=m^2\phi^2/2$. 
The vacuum energy is given by the minimum of the effective potential
$V(\phi)$; since for a free scalar field the 1-loop contribution $V_1$ 
is $\phi$-independent and there are no higher-loop contributions, the
vacuum energy just coincides with $V_1$. 

We shall now show, however, that the differentiation with respect to $m^2$ 
that has been used in the derivation of (\ref{rho3}) is in general ambiguous 
and may lead to incorrect results. Comparing the integrands of the right-hand 
sides of eqs. (\ref{rho1}) and (\ref{press1}) and assuming that the 
differentiation with respect to $m^2$ is justified, one finds 
\be
\left(1-2m^2 \frac{\partial}{\partial m^2}\right)\rho=(d-1) p\,. 
\label{rel2}
\ee
Here we have introduced the factor $d-1$ on the r.h.s. so that 
the relation is suitable also for the dimensionally-continued case. 
It is easy to see that in the case of the dimensional regularization the 
right-hand sides of eqs. (\ref{rho2}) and (\ref{press2}) indeed satisfy 
(\ref{rel2}), and the relativistic invariance condition (\ref{rel1}) is 
also fulfilled. At the same time, if one uses the Euclidean cutoff as a 
regulator (in conjunction with the differentiation with respect to $m^2$),
\mbox{eq.~(\ref{rel1})} is not satisfied. This can be most directly 
seen by substituting the explicit formula (\ref{rho4}) for $\rho$ into 
(\ref{rel2}) and computing the pressure $p$. 
Another way to see this is to assume that eq. (\ref{rel1}) is satisfied; 
from (\ref{rel2}) one then obtains 
\be
m^2\frac{\partial \rho}{\partial m^2} = \frac{d}{2}\, \rho\,,
\label{rel3}
\ee
which in $d=4$ dimensions has the unique solution $\rho=const\cdot m^4$, 
in clear conflict with eq. (\ref{rho4}). Thus, eq. (\ref{rho4}) cannot be 
trusted.

The failure of the approach based on the differentiation with respect to 
$m^2$ is a consequence of the fact that momentum cutoffs (either 4-dimensional 
or 3-dimensional) violate one of the two conditions (i) and (ii) discussed 
above, so that this differentiation is in fact illegitimate. On the other
hand, in the case of dimensional regularization both conditions are satisfied; 
eq. (\ref{rel3}) yields $\rho=c_1 \mu^4 (m^2/\mu^2)^{d/2}$ 
with $\mu$ an arbitrary mass parameter and $c_1$ a ($d$-dependent)  
dimensionless constant. This is in accord with eq. (\ref{rho2}). 
Applying dimensional regularization directly to (\ref{rho3}) one recovers 
eq. (\ref{rho2}). 

Instead of using the differentiation with respect to $m^2$ in order to 
arrive at relativistically invariant \mbox{expressions} for the vacuum 
energy
and pressure, one could have \mbox{started} from the effective potential 
as derived 
from the functional integral formalism. Since this formalism is manifestly 
covariant, condition (\ref{rel1}) will be automatically satisfied.
However, the functional integral \mbox{formalism} yields the effective 
potential
only up to a constant (field-independent) term. While this is quite 
sufficient when one is looking for the vacuum configuration of the 
\mbox{fields}, 
it is not suitable for our purposes since this field-independent constant 
is exactly what we are seeking.  We therefore adopt a different approach.

Let us calculate the trace of the energy momentum tensor of the field
$\phi$. Using the equation of motion, one formally finds
\be
T^\mu_\mu = m^2 \phi^2\,. 
\label{tr1}
\ee
%
In the case of interacting fields this result is modified by quantum
anomalies \cite{TA}, but for the free-field case under consideration 
it is not anomalous in flat spacetimes. In curved spacetimes, the trace 
of the energy-momentum tensor is modified by the conformal anomaly \cite{CA}. 
This adds finite terms to the r.h.s. of (\ref{tr1}) and thus does not affect 
the character of the divergences of the vacuum energy; we therefore ignore 
these terms.

{}From eq. (\ref{tr1}), the vacuum expectation value of $T^\mu_\mu$ is 
$\langle 
0|T^\mu_\mu|0\rangle=m^2 \langle 0|\phi^2|0\rangle$.
On the other hand, from (\ref{rel1}) one finds 
$\langle 0|T^\mu_\mu|0\rangle =\rho-3p=4\rho$. 
This gives 
\be
\rho = \frac{1}{4} m^2 \,\langle 0|\phi^2|0\rangle\,.
\label{rho6} 
\ee
The vacuum expectation value of $\phi^2$ has a well known integral 
representation, and one finally finds
\be
\rho=\frac{m^2}{4}\int\frac{d^4k}{(2\pi)^4}
\frac{i}{k^2-m^2+i\varepsilon}=\frac{m^2}{4}D_F(x,x)\,.
\label{rho7}
\ee
The integral here coincides with the one in eq. (\ref{drho2}); however,  
in the derivation of eq. (\ref{rho7}) no differentiation with respect to
$m^2$ has been used. 
By direct power counting, the leading divergence of the integral in eq. 
(\ref{rho7}) is quadratic. 
Performing Wick rotation and imposing the Euclidean 
4-momentum cutoff one finds 
\be
\rho=\frac{1}{64\pi^2}\left[\Lambda^2 m^2 - m^4\ln\left(
\frac{\Lambda^2+m^2}{m^2}\right)\right]\,. 
\label{rho8}
\ee
A similar result is obtained if one uses the Pauli-Villars regularization 
rather than the 4-momentum cutoff \cite{ff}. 
An important consequence of the absence of the quartic divergence 
in the zero-point energy  is that it vanishes for massless 
fields.

Several comments are in order. 

(1) The unrenormalized vacuum energy density $\rho$ that we considered is 
obviously regularization scheme dependent; therefore the right question to
ask is not if one or another scheme results in a quartic divergence, but 
rather if there exists {\it any} relativistically invariant regularization 
that leads to ${\cal O}(\Lambda^4)$ terms in $\rho$. 
The fact that several regularization schemes leading to quartic 
divergences in eqs. (\ref{rho1}) and (\ref{press1}) violate the 
relativistic invariance condition (\ref{rel1}) does not answer this 
question because 
it does not rule out the existence of a such a consistent, relativistically
invariant regularization. However, from the representation (\ref{rho6}) for 
the zero-point energy it directly follows by a simple power counting that 
$\rho$ cannot have quartic divergences.

(2) In our discussion we considered flat spacetimes. In a curved    
spacetime, the curvature only influences the modes with the wavelength
$k^{-1}\aprge a$ where $a$ is the local curvature radius, as one can
always find a local Lorentz frame in which the spacetime is flat in the
region of the size $r\aprle a$. Since the ultraviolet divergences
correspond to the smallest wavelengths with $k^{-1}\ll a$, their nature
does not depend on whether the spacetime is flat  or curved.
Therefore the divergence of the vacuum energy is unaffected by the
spacetime curvature.

(3) In models with broken supersymmetry the vacuum energy is known to 
have no quartic divergence as a result of the cancellations between the 
\mbox{contributions} of the fermionic and bosonic
superpartners. We have shown here that in fact the quartic divergences
are absent even without supersymmetry since the relativistic invariance 
precludes their appearance in the {\em individual} \mbox{contributions} of   
different fields. 

(4) We have considered the zero-point energies of free fields. As was 
pointed out above, in the interacting case the trace formula (\ref{tr1}) is 
modified by the flat-space trace anomaly \cite{TA}; therefore this case
requires a special study. 

(5) Although our results have not led to a solution of the cosmological   
constant problem, we believe that they can be relevant to it. 
{}From the absence of the quartic divergences it follows that in flat 
spacetimes the zero-point energies of free massless fields vanish. 
In particular, the zero-point energy of the free electromagnetic  
field is zero \cite{f2}. In a curved spacetime of curvature radius $a$ the 
vacuum energy of a free massless field is of order $a^{-4}$; for the
present-day \mbox{universe} this is vanishingly small compared to the 
critical energy density $\rho_c$. Thus, free massless fields give
negligible contributions to the vacuum energy.

{\em Acknowledgements.} The author is grateful to M.C. Bento, Z. Berezhiani, 
H.B. Nielsen, L.B. Okun, M.E. Shaposhnikov and A.Yu. Smirnov for useful
discussions, and
to V.A. Rubakov and J. Smit for useful correspondence. He was supported by
the Calouste Gulbenkian Foundation as a Gulbenkian Visiting Professor at 
Instituto Superior T\'ecnico.



\begin{references}
\bibitem{ct} 
S.~Perlmutter {\it et al.}  
Astrophys.\ J.\  {\bf 517}, 565 (1999);
%
A.~G.~Riess {\it et al.}  
Astron.\ J.\  {\bf 116}, 1009 (1998).

\bibitem{rev} For reviews see, e.g., 
%
S.~Weinberg, 
Rev.\ Mod.\ Phys.\  {\bf 61}, 1 (1989); 
%
V.~Sahni and A.~Starobinsky,
Int.\ J.\ Mod.\ Phys.\ D {\bf 9}, 373 (2000);
%
S.~M.~Carroll,
Living Rev.\ Rel.\  {\bf 4}, 1 (2001).

\bibitem{f1} There are several ways to arrive at this result. One may 
note that the vacuum can be considered a perfect fluid; using the general 
form of the energy-momentum tensor of perfect fluids $T_{\mu\nu}=-p g_{\mu\nu}
+(\rho+p)U_\mu U_\nu$ where $U_\mu$ is the four-velocity vector, one
obtains (\ref{rel1}). Alternatively, one may start from a flat spacetime with 
Minkowski metric $g_{\mu\nu}=\eta_{\mu\nu}\equiv diag(1, -1, -1, -1)$ so that 
eq. (\ref{rel1}) is automatically satisfied, and then notice that any curved 
spacetime is locally flat.  



\bibitem{TA}
C.~G.~Callan, S.~R.~Coleman and R.~Jackiw,
Annals Phys.\  {\bf 59}, 42 (1970).
%

\bibitem{CA}
D.~M.~Capper and M.~J.~Duff,
Nuovo Cim.\ A {\bf 23}, 173 (1974);
%
S.~Deser, M.~J.~Duff and C.~J.~Isham,
Nucl.\ Phys.\ B {\bf 111}, 45 (1976).

\bibitem{ff}
Note that the Pauli-Villars regularization cannot be applied directly 
to eq. (\ref{rho1}) because when the regularization is lifted (i.e. in the 
limit of the infinite masses of the regulator fields) the original 
divergent expression is not recovered in that case. 


\bibitem{f2}
This does not apply to the zero-point energies inside
cavities or when the boundaries are present, and so the Casimir energies do
not vanish, of course. However, when calculated in a covariant way and
with a relativistically invariant regularization, these energies must be
finite, and no subtraction of the energy of the free vacuum is necessary.

\end{references}
\end{document}